\documentclass[pra,aps,notitlepage,superscriptaddress,showpacs,showkeys]{revtex4-1}
\usepackage{amsmath,amsthm,amssymb,graphicx,subfigure,xcolor}
\usepackage[unicode=true]{hyperref}
\hypersetup{
     colorlinks=true,       		
     linkcolor=blue,          	
     citecolor=red,            
     urlcolor=magenta,           	
 }
\newcommand{\ket}[1]{|#1\rangle}

\newcommand{\ketbra}[2]{|#1\rangle\langle#2|}
\newcommand{\tr}{{\rm tr}}

\begin{document}

\title{Monte Carlo approach to the evaluation of the security of device-independent quantum key distribution}

\author{Hong-Yi~Su}
\email{hysu@hrbeu.edu.cn}
\affiliation{Key Laboratory of In-Fiber Integrated Optics, Ministry of Education, College of Physics and Optoelectronic Engineering, Harbin Engineering University, Harbin 150001, People's Republic of China}
\affiliation{Key Laboratory of Photonic Materials and Device Physics for Oceanic Applications, Ministry of Industry and Information Technology of China, College of Physics and Optoelectronic Engineering, Harbin Engineering University, Harbin 150001, People's Republic of China}

\date{\today}
\begin{abstract}
We present a generic study on the information-theoretic security of multi-setting device-independent quantum key distribution protocols, i.e., ones that involve more than two measurements (or inputs) for each party to perform, and yield dichotomic results (or outputs). The approach we develop, when applied in protocols with either symmetric or asymmetric Bell experiments, yields nontrivial upper bounds on the secure key rates, along with the detection efficiencies required upon the measuring devices. The results imply that increasing the number of measurements may lower the detection efficiency required by the security criterion. The improvement, however, depends on (i) the choice of multi-setting Bell inequalities chosen to be tested in a protocol, and (ii) either a symmetric or asymmetric Bell experiment is considered. Our results serve as an advance toward the quest for evaluating security and reducing efficiency requirement of applying device-independent quantum key distribution in scenarios without heralding.
\end{abstract}

\maketitle

\section{Introduction}
Device-independent quantum key distribution (DIQKD)~\cite{Ekert91,MY98,BHK05,AGM06,AMP06,ABG+07,PAB+09,PAM+10,MPA11,AMP12,TH13,VV14,MS16,A-FDF+18,A-FRV19,MvDR+19} requires minimal assumptions on implementation devices to certify an information-theoretic security, thus serving as one of promising ways to ruling out side-channel attacks and narrowing the ideal-versus-practical gap in security. It involves Bell tests (BT)~\cite{BCP+14} accomplished in a loophole-free manner~\cite{HBD+15,GVW+15,SM-SC+15,RBG+17,LWZ+18}, to quantitatively evaluate  eavesdropping by some potential adversary, Eve. Both DIQKD and BT, however, suffer compromised consequences without closing the loopholes~\cite{Bell04,Pearle70,CH74,GG99}. To the end, one approach, among others, to closing the detection loophole, one of the principal loopholes in BT, resorts to increasing the number of measurements~\cite{Massar02,MPR+02,MP03,BG08,PV09,VPB10,Branciard11}. It has been estimated that performing more measurements in each party may provide improvement in decreasing the threshold efficiency. Namely, one can have $\eta_{\min}\geq (m_A+m_B-2)/(m_A m_B-1)$~\cite{MP03}, where $m_{A,B}$ are the numbers of measurements (settings) of the two parties, Alice and Bob. One then is led to be expecting that, to get shared secure keys, increasing the number of measurements in DIQKD, where the detection loophole is more serious than in BT, may stand to benefit from the decrease of the threshold efficiency in BT.

Several difficulties with investigating multi-setting DIQKD protocols arise and---in spite of a few case studies---remain. First, the general quantity $\eta_{\min}$ for efficiencies in~\cite{MP03} represents a lower bound; whether it can always be reached, or to what extent it can be adapted to cryptographic protocols, is yet to be conclusive. Second, due to the particular structure of the Clauser-Horne-Shimony-Holt (CHSH) inequality~\cite{Bell64,CHS+69}, it suffices to consider a $2\times2$-dimensional Hilbert subspace in the security analysis of the CHSH-based protocol~\cite{ABG+07,PAB+09}. Multi-setting protocols do not share such a mathematical structure, rendering the security analysis all the more complicated. In fact, without affecting the faithfulness in experiment, there have been a number of methods to improving the efficiencies by having undesired data deducted. For instance, one can use the heralding method (see, e.g.,~\cite{ZLA-F+23} for the concepts and references therein; also~\cite{XMZ+20,PR22,PGT+23}), e.g., via the non-destructive detection~\cite{NFL+21} or the qubit amplifiers~\cite{ZC19,GPS10,PMW+11,CM11,M-SBB+13,KMS+20}, to select signals to trigger detection, while safely neglecting undetected events; or, one can perform a classical data post-processing scheme for the error-correction purpose to improve the key rates~\cite{ML12}.

Without heralding, to reduce the efficiency requirement in DIQKD still remains an important question. The following two efficiencies of measuring devices are often the cases in practice~\cite{BGS+07}:
\begin{equation}\nonumber
\begin{split}
  &{\rm symmetric~detection~setup:}~~~~\eta_A=\eta_B=\eta,\\
  &{\rm asymmetric~detection~setup:}~~~~\eta_A=1,~\eta_B=\eta,
\end{split}
\end{equation}
where $\eta_{A,B}$ are efficiencies of Alice's and Bob's measuring devices, respectively. The symmetric case denotes equally non-ideal measurements by Alice and Bob when the information carriers are of the same type of particles (e.g., entangled photon pairs); the asymmetric case denotes ideal-versus-non-ideal measurements when they are not of the same type of particles (e.g., entangled atom-photon pairs~\cite{BMD+04,MMB+04,VWS+06} with which Alice measures the atoms and Bob measures the photons). In general, it is a simple way to come up with a DIQKD protocol out of the BT by adding to Bob one more measurement which coincides with one of Alice's measurements along, say, the $\hat z$-axis, and using this pair of measurements to generate keys.

In this paper, we develop a Monte Carlo approach to evaluating upper bounds on the secure key rates in multi-setting two-output DIQKD protocols against collective attacks. We consider both symmetric and asymmetric detection setups in BT and key generation. The threshold efficiency of the CHSH-based protocol is found to be approximately $85.8\%$ in the asymmetric setup, less than the well-known $92.4\%$ in the symmetric setup~\cite{PAB+09}. The threshold efficiency of the $I_{3322}$-based protocol is estimated to be approximately $83.6\%$ in the asymmetric setup, as well as those of a large family of multi-setting-inequality-based protocols being estimated to be less than $92.4\%$, too. A remark here is that it is the modeling of undetected events presented in~\cite{PAB+09} that we take to use in this paper to evaluate the eavesdropping information. There have been literatures wherein different modelings are used (for instance, see~\cite{BFF21,WAP21,BFF21-2,MPW22,LB-JF+22} for recent methods to evaluating the eavesdropping). In general, taking account of more input-output statistics of undetected events, e.g., adding one more output to each party to register the undetected event, may yield slightly higher key rates. The key rates can be further improved by considering partially entangled states and noisy-preprocessing techniques (see, e.g.,~\cite{LB-JF+22}), thus the resulting threshold efficiencies may be improved as well.

The paper is structured as follows. We first describe the general setup for DIQKD protocols with undetected events taken into account. We then present the evaluation of key rates by developing a formalism in which the mutual information and eavesdropping information are put in a same index. It is followed by an explicit procedure of \textcolor[rgb]{0.00,0.50,1.00}{a} Monte Carlo approach, along with case studies of the CHSH- and $I_{3322}$-based~\cite{CG04} protocols and the BB84-type protocol~\cite{BB84}. The efficiencies and critical error rates are estimated with Bell diagonal states. More protocols with other multi-setting Bell inequalities are also studied. We proceed beyond the $2\times2$-dimensional subspaces and generalize the Monte Carlo approach to arbitrarily high-dimensional systems. The paper ends with a summary and outlook.

\section{Security of multi-setting quantum key distribution}

\subsection{General setup}
Let us assume the measurements used in the security proof be projective ones. To justify the use, note that if one uses the positive-operator-valued measures (POVMs) to perform measurements, they can readily be transformed to projectors by the Naimark theorem. Obviously, projective measurements become POVMs when undetected events are taken into account (see, e.g., Eq.~(\ref{efficiency:basic}) below ). In general scenarios, the projectors are not necessary to be of rank one. By requiring that measurement must produce, for instance, two distinct outcomes, the projectors could be of rank two or higher (see, e.g., Eq.~(\ref{proj:d-3}) below).

Consider bipartite Hilbert spaces of dimension $d\times d$ and a measurement set $M=\{ 1,2,...,m \}$. Given projectors
\begin{equation}
\begin{split}
  \forall~i,j\in M,~u_i,v_j\in SU(d):
  ~~~~\{ \hat{p}^{(a)}_{u_i}|a=0,1,...,d-1 \},~~~
  \{ \hat{p}^{(b)}_{v_j}|b=0,1,...,d-1 \},
  \end{split}
\end{equation}
respectively for Alice and Bob, the joint correlation is computed as $P(a,b|i,j)=\tr (\hat{p}^{(a)}_{u_i}\otimes \hat{p}^{(b)}_{v_j} \rho_{AB})$ quantum mechanically, where Alice chooses her $i$-th measurement, getting result $a$, Bob chooses his $j$-th measurement, getting result $b$, $\rho_{AB}$ denotes the state shared by Alice and Bob, and $u_i,v_j$ are unitary transformations in $SU(d)$ group. Let us tacitly assume, when $u_i,v_j$ take unit identity $\openone$, that one can have $\hat{p}^{(a)}_{\openone}=\ketbra{a}{a}$ and $\hat{p}^{(b)}_{\openone}=\ketbra{b}{b}$ where $\{ \ket{k}|k=0,1,...,d-1 \}$ span a set of computational bases.

An $m$-setting $d$-dimensional Bell inequality, denoted by $I_{mmdd}$, can be expressed as
\begin{equation}
  \mathsf{P}=\sum_{i,j}\sum_{a,b}\omega^{ab}_{ij}P(a,b|i,j)\leq0,~~~~\forall i,j\in M,
\end{equation}
with $\omega^{ab}_{ij}$'s coefficients properly chosen to make the inequality nontrivial, i.e., violated in quantum mechanics. The inequality holds for all local-hidden variable models~\cite{Bell64}. Let us denote quantum violation as
\begin{equation}
  \mathsf{Q}=\max\Big\{ \mathsf{P} \Big|i,j\in M,~u_i,v_j\in SU(d) \Big\}>0 \label{QM}
\end{equation}
with respect to some $\rho_{AB}$. It should be noted that a large family of Bell inequalities reach their maximum quantum violations under very particular sets of measurements (e.g., Bell multi-port measurements~\cite{ZZH97}) but for the sake of generality we maintain $u_i,v_j$ as arbitrary unitary transformations.

To model undetected events in experiment outputting dichotomic results $0$ and $1$, let us assign the undetected with $0$, and so with an $\eta$ detection efficiency the projectors become non-ideal, reading actually
\begin{align}
  \hat{p}^{(0)}_\tau\overset{\eta}{\longrightarrow}\eta\times\hat{p}^{(0)}_\tau+(1-\eta),~~~~\hat{p}^{(1)}_\tau\overset{\eta}{\longrightarrow}\eta\times\hat{p}^{(1)}_\tau,\label{efficiency:basic}
\end{align}
with $\tau=u_i,v_j$. The term $(1-\eta)$ is due to the undetected events $0$ and $1$, both of which happen with probability $1-\eta$ and according to the model should then be assigned $0$, contributing to the actual $\hat{p}^{(0)}$. Applying (\ref{efficiency:basic}) to joint correlations $P(a,b|i,j)$ of Alice and Bob yields (see also~\cite{G-UPC21})
\begin{equation}
\begin{split}
  P(a,b|i,j)\longrightarrow
  \eta_A\eta_B P(a,b|i,j)+\delta_{a,0}(1-\eta_A)\eta_B P_B(b|j)+\delta_{b,0}\eta_A(1-\eta_B) P_A(a|i)+\delta_{a,0}\delta_{b,0}(1-\eta_A)(1-\eta_B),
\end{split}
\end{equation}
where $\delta$ denotes the Kronecker function, and $P_{A,B}$ are marginals.

\subsection{\label{keyrate}Evaluation of key rates}
We choose to take the information-theoretic key rate to analyze the security. One needs to first evaluate the eavesdropping information $I_E$ and the mutual information $I_{AB}$, and then attain in the infinite-length limit the key rate $\mathcal{R}=I_{AB}-I_E$~\cite{DW05}. Let us stress, however, that the use of Bell diagonal states in what follows is a strong assumption on the security, and that in our approach finding optimal states that maximize Eve’s information may not be guaranteed for more-than-two-input scenarios in general. Hence the $\mathcal{R}$'s evaluated below serves as upper bounds on the key rates.

To compute $I_E$, one needs to optimize the Holevo quantity $\chi=S(\rho_E)-\left[S(\rho_{E|0})+S(\rho_{E|1})\right]\big/2$, where $S(\rho)$ is the von Neumann entropy, $\rho_E$ denotes Eve's reduced state, and $\rho_{E|a}$ denotes Eve's normalized conditional state upon Alice's measurement $\hat p_u^{(a)}$. Since there exists an overall distilled pure state $\ket{\psi}_{ABE}$ from which $\rho_E$ and $\rho_{AB}$ can be attained in a reduction way, one has $S(\rho_E)=S(\rho_{AB})$. The fact that instead of all possible bipartite states it suffices to consider the Bell-diagonal state~\cite{ABG+07,PAB+09}, reading $\rho_{AB}=\sum_{i=0}^3 \Lambda_i \ketbra{\Phi_i}{\Phi_i}$ with $\ket{\Phi_0}=(\ket{00}+\ket{11})/\sqrt{2}$, $\ket{\Phi_1}=(\ket{00}-\ket{11})/\sqrt{2}$, $\ket{\Phi_2}=(\ket{01}+\ket{10})/\sqrt{2}$, $\ket{\Phi_3}=(\ket{01}-\ket{10})/\sqrt{2}$, and $\sum_i\Lambda_i=1$, leads to $S(\rho_{E|0})=S(\rho_{E|1})$.

To get further simplification, note that $\rho_{E|a}$ is of rank two, with eigenvalues (which are derived in the $\hat{x}\hat{z}$-plane in~\cite{PAB+09}; see also~\cite{Su22})
\begin{equation}
\begin{split}
  \Lambda_\pm=\frac{1}{2}\biggr[ 1\pm\big[(\mu_+ -\nu_+)^2\cos^2\theta+
  (\mu_-^2+\nu_-^2+2\mu_-\nu_-\cos2\phi)\sin^2\theta\big]^{1/2}  \biggr],
\end{split}
\end{equation}
with respect to Alice's measurements $\hat p_u^{(0,1)}=\ketbra{\pm \hat n}{\pm \hat n}$,
where $\mu_\pm=\Lambda_0\pm\Lambda_1$, $\nu_\pm=\Lambda_2\pm\Lambda_3$, and
\begin{equation}
  \begin{split}
    \ket{+\hat n}=\cos\frac{\theta}{2}\ket{0}+\sin\frac{\theta}{2}e^{i\phi}\ket{1},\\
    \ket{-\hat n}=\sin\frac{\theta}{2}\ket{0}-\cos\frac{\theta}{2}e^{i\phi}\ket{1}.
  \end{split}
\end{equation}
With all these combined, the Holevo quantity becomes $\chi=H(\vec\Lambda)-h(\Lambda_\pm)$, with the classical Shannon entropy $H(\vec x):=-\sum_i x_i\log_2 x_i$ and the binary entropy $h(y):=-y\log_2 y-(1-y)\log_2(1-y)$.

Because the measurements may not be faithfully performed, the Holevo quantity should be presumed to be maximized, or equivalently, taking into account the monotonicity of $h(y)$, the $\Lambda_-$ (or $\Lambda_+$) should be presume to be minimized (or maximized). Let
\begin{equation}
  {\partial\Lambda_\pm}/{\partial \theta}={\partial\Lambda_\pm}/{\partial \phi}=0,
\end{equation}
and we find
\begin{equation}
  \theta,\varphi\in\{ 0,~\pi/2,~\pi \}.
\end{equation}
It follows that $\chi$ has three extrema:
\begin{equation}
\chi^{(k)}=H(\vec \Lambda)-h(\Lambda_\pm^{(k)}),~~~~~~j=1,2,3,
\end{equation}
where
\begin{equation}
\begin{split}
&\Lambda_\pm^{(1)}=\frac{1}{2}\Big[ 1\pm | \Lambda_0+\Lambda_1-\Lambda_2-\Lambda_3 | \Big],\\
&\Lambda_\pm^{(2)}=\frac{1}{2}\Big[ 1\pm | \Lambda_0-\Lambda_1+\Lambda_2-\Lambda_3 | \Big],\\
&\Lambda_\pm^{(3)}=\frac{1}{2}\Big[ 1\pm | \Lambda_0-\Lambda_1-\Lambda_2+\Lambda_3 | \Big].
\end{split}
\end{equation}
Let $\mathbb{B}$ be the set of all parameters in Bell-diagonal states, and $\mathbb{B}_\mathsf{Q}\subseteq\mathbb{B}$ be a subset that corresponds to some particular value of quantum violation $\mathsf{Q}$. The eavesdropping information is found to be
\begin{equation}
\forall~\mathsf{Q}:~~~~ I_E=\max\Big\{\chi^{(k)}\Big|k=1,2,3,~\vec\Lambda\in\mathbb{B}_\mathsf{Q}\Big\}.\label{eaves:evaluate}
\end{equation}
In other words, one can thus have a $\mathsf{Q}$-dependent $I_E$.

To compute $I_{AB}$, one needs first to get the quantum bit error rate (QBER) $\varepsilon:=\tr [(\hat p_{\openone}^{(0)}\otimes\hat p_{\openone}^{(1)}+\hat p_{\openone}^{(1)}\otimes\hat p_{\openone}^{(0)} ) \rho_{AB}]$. Taking undetected events into account, it becomes
\begin{equation}
  \varepsilon\longrightarrow\eta_A\eta_B~\varepsilon+\Big[\eta_A(1-\eta_B)+\eta_B(1-\eta_A)\Big]\Big/2,
\end{equation}
with $\varepsilon=\Lambda_2+\Lambda_3$ for the Bell-diagonal state. It is well acceptable that one can assume $\Lambda_1=\Lambda_2$ due to the preparation of states in practice, getting the QBER equal in the $\hat z$- and $\hat x$-axes. Hence one gets the mutual entropy $ \Gamma=1-h(\varepsilon)$. Despite that the $\varepsilon$, or $\Gamma$, in general, is irrelevant to $\mathsf{Q}$, it is vital in our approach that one must define a $\mathsf{Q}$-dependent mutual information, in order to evaluate the key rate.

To the end, let $\mathbb{B}_\Gamma$ be a subset that corresponds to some particular value of mutual entropy $\Gamma$. As such, the states in $\mathbb{B}_\Gamma$ will yield a domain of quantum violation, namely,
\begin{equation}
\forall~\Gamma:~~~~\vec\Lambda\in\mathbb{B}_\Gamma\longmapsto\mathsf{Q}\in[\mathsf{Q}^{\min}_\Gamma,~\mathsf{Q}^{\max}_\Gamma].
\end{equation}
Then some $\mathsf{Q}_\Gamma$ must exist in the domain such that $I_E$ reaches a local maximum:
\begin{equation}
\begin{split}
\exists~\mathsf{Q}_\Gamma:~~~~
 I_E(\mathsf{Q}_\Gamma)=\max\Big\{ I_E(\mathsf{Q}) \Big| \mathsf{Q}\in[\mathsf{Q}^{\min}_\Gamma,~\mathsf{Q}^{\max}_\Gamma] \Big\}.
\end{split}
\end{equation}
For all pairs $(\mathsf{Q}_\Gamma,\Gamma)$, there must be a mapping, namely,
\begin{equation}
\mathsf{define}~~~~~~I_{AB}:~~\mathsf{Q}_\Gamma\longmapsto\Gamma,\label{mutual}
\end{equation}
 which can thus be used as a $\mathsf{Q}$-dependent $I_{AB}$.

It is here remarked, for one thing, that in computing $I_{AB}$, non-ideal efficiencies (i.e., undetected events) should be taken into account; but in computing $I_E$ the efficiencies should be presumed ideal, because Eve gets the most information when Alice faithfully performs measurements in order to try with Bob to generate keys. For another, as derived above, one can have both $I_E$ and $I_{AB}$ dependent on $\mathsf{Q}$; however, the state $\vec\Lambda\in\mathbb{B}$ that yields $I_E(\mathsf{Q})$ is not required to be the same as the state $\vec\Lambda'\in\mathbb{B}$ that yields $I_{AB}(\mathsf{Q})$. If, instead, one requires the states that yield $\mathsf{Q}$ to be the same in computing $I_E(\mathsf{Q})$ and $I_{AB}(\mathsf{Q})$, as well as specifying Alice's measurements in evaluating Eve's conditional states (i.e., replacing the maximization (\ref{eaves:evaluate}) with a Holevo quantity that corresponds to that Alice always measuring along the $\hat z$-axis), then the protocol reduces to a BB84 protocol in which the device-independent feature is lost (see the example below; also \cite{ABG+07,PAB+09}).

\subsection{Procedure of the Monte Carlo approach}
The procedure is summarized in Table.~\ref{table1}. Once the key rate is attained, it is immediate to derive a number of critical values, e.g., the maximum QBER and the threshold efficiency for a selected protocol.

\begin{table}[t]
\caption{\label{table1}Monte Carlo procedure to evaluating an upper bound on the secure key rate in the DIQKD protocol.}
\begin{ruledtabular}
\begin{tabular}{cp{165mm}}
 Steps & Descriptions for each step\\
 \hline
  1 & Select a DIQKD protocol in which the Bell test and key generation are specified.\\
  2 & Randomly generate a Bell-diagonal state $\rho_{AB}$ and compute the Holevo quantity $\chi=\max\{\chi^{(k)}|k=1,2,3\}$.\\
  3 & Compute the quantum violation $\mathsf{Q}$ of the Bell inequality $\mathsf{P}$ with $\rho_{AB}$.\\
  4 & Record the pair $(\mathsf{Q},\chi)$.\\
  5 & Repeat steps 1 through 4 sufficiently many times to collect the data of pairs $(\mathsf{Q},\chi)$, the upper bound of which corresponds then to the eavesdropping information $I_E(\mathsf{Q})$.\\
  6 & Randomly generate a Bell-diagonal state $\rho'_{AB}$ and compute the QBER $\varepsilon$, from which to get the mutual entropy $\Gamma$.\\
  7 & Compute the quantum violation $\mathsf{Q}$ of the Bell inequality $\mathsf{P}$ with $\rho'_{AB}$.\\
  8 & Record the pair $(\mathsf{Q},\Gamma)$.\\
  9 & Repeat steps 6 through 8 sufficiently many times to collect the data of pairs  $(\mathsf{Q},\Gamma)$, from which, together with the definition (\ref{mutual}), one can determine the mutual information $I_{AB}(\mathsf{Q})$.\\
  10 & Evaluate the key rate $\mathcal{R}=I_{AB}(\mathsf{Q})-I_E(\mathsf{Q})$ with respect to any $\mathsf{Q}$.
\end{tabular}
\end{ruledtabular}
\end{table}

\subsection{Case studies}
We illustrate our approach in two examples, alongside a reduction to the BB84 protocol.

\subsubsection{The CHSH-based protocol}
It suffices here to consider Bell-diagonal states, since it has been proved that $2\times2$ dimensional subspaces can always be found in evaluating the key rate for any two-input-two-output scenarios~\cite{PAB+09}. The CHSH inequality reads $\mathsf{P}=P_{11}+P_{12}+P_{21}-P_{22}-P_{10}-P_{01}\leq0$ (which is actually the Clauser-Horne inequality~\cite{CH74} in mathematical equivalence), with short notations $P_{ij}:=P(0,0|i,j)$, $P_{0j}:=\sum_a P(a,0|i,j)$, and $P_{i0}:=\sum_b P(0,b|i,j)$.

\begin{figure}[t]
\subfigure[]{\includegraphics[width=40mm]{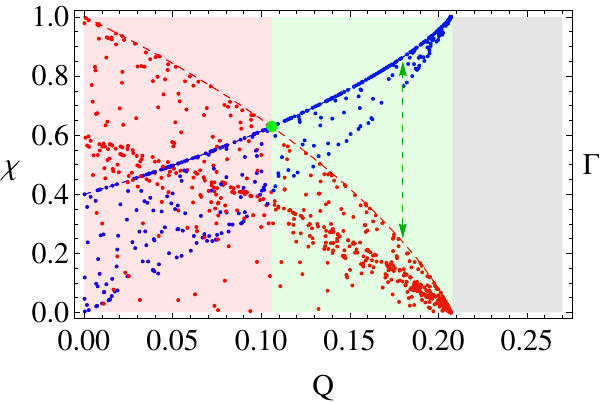}}\hspace{4mm}
\subfigure[]{\includegraphics[width=40mm]{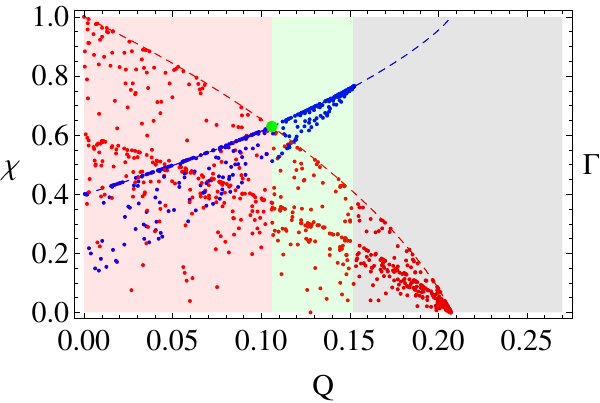}}\hspace{4mm}
\subfigure[]{\includegraphics[width=40mm]{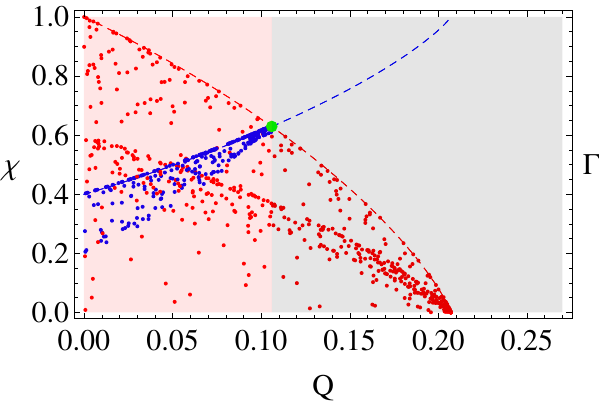}}\hspace{4mm}
\subfigure[]{\includegraphics[width=40mm]{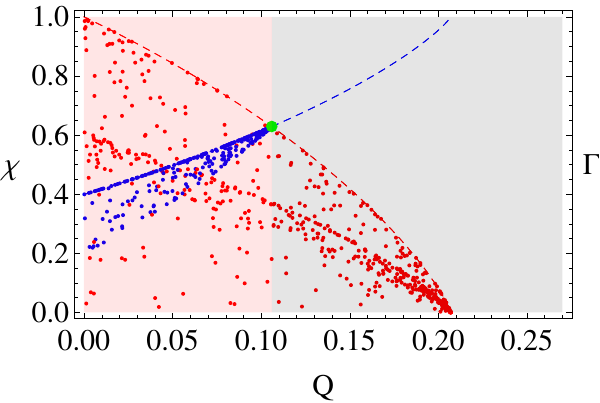}}\\
\subfigure[]{\includegraphics[width=40mm]{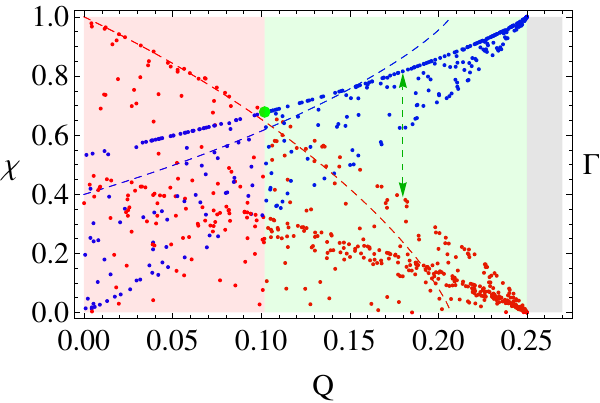}}\hspace{4mm}
\subfigure[]{\includegraphics[width=40mm]{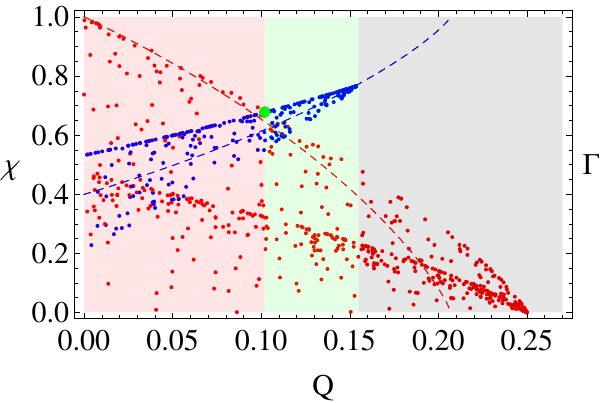}}\hspace{4mm}
\subfigure[]{\includegraphics[width=40mm]{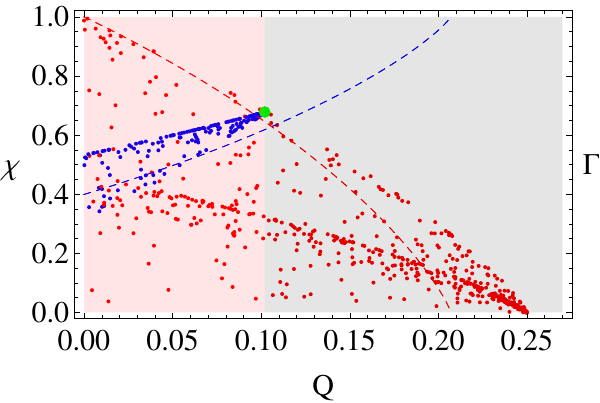}}\hspace{4mm}
\subfigure[]{\includegraphics[width=40mm]{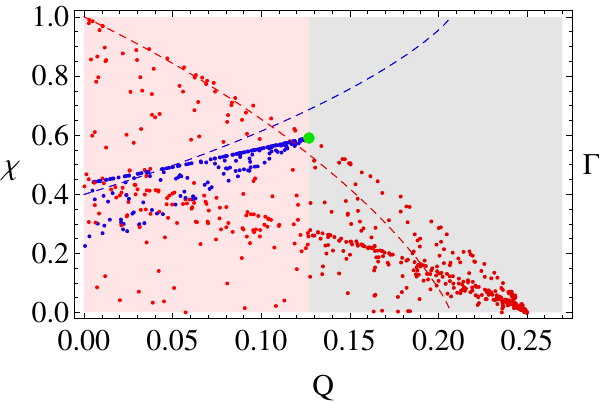}}
  \caption{(Color online) The $\chi$ and $\Gamma$ figures with respect to $\mathsf{Q}$ under various detection efficiencies in the CHSH- and $I_{3322}$-based protocols. The red points represent the data of $(\mathsf{Q},\chi)$, the upper boundary of which pins down the $I_E(\mathsf{Q})$. The blue points represent the data of $(\mathsf{Q},\Gamma)$, the upper boundary of which pins down the $I_{AB}(\mathsf{Q})$ (see the main text for the reason). For the CHSH-based protocol: in (a), (b), and (c), we take the symmetric setup with $\eta=100\%,~96\%,~92.4\%$, respectively; in (d), we take the asymmetric setup with $\eta=85.8\%$. For the $I_{3322}$-based protocol: in (e), (f), and (g), we take the symmetric setup with $\eta=100\%,~96\%,~93.8\%$, respectively; in (h), we take the asymmetric setup with $\eta=83.6\%$.  The dashed curves in the last four figures are analytic results of the CHSH-based protocol, plotted for the sake of comparison. }\label{fig1}
\end{figure}

Following the Monte Carlo procedure, we pin down $I_E(\mathsf{Q})$ and $I_{AB}(\mathsf{Q})$, as shown in Figs.~\ref{fig1}(a) through \ref{fig1}(d). The argument is as follows. First, it is immediate to see from the figure that for each $\mathsf{Q}$ there corresponds to non-unique $\chi$, and the $I_E(\mathsf{Q})$ serves as the upper bound. Then, for each $\Gamma$ there corresponds to non-unique $\mathsf{Q}$ and, according to the definition (\ref{mutual}), the $I_{AB}(\mathsf{Q})$ should be taken at a value of $\mathsf{Q}$ at which $I_E(\mathsf{Q})$ reaches a local maximum. In the CHSH case, $I_E$ always decreases with the increase of $\mathsf{Q}$, meaning that for each $\Gamma$, it is the $\mathsf{Q}_{\Gamma}^{\min}$ that yields a maximum $I_E$. The $I_{AB}(\mathsf{Q})$ is henceforth pinned down.

The critical QBER corresponds to the cross point
\begin{equation}
  (\mathsf{Q},I_E)=(\mathsf{Q},I_{AB})\simeq(0.106, 0.629),
\end{equation}
from which we find $\varepsilon_{\rm cr}\simeq7.15\%$. The threshold efficiency can be attained when there is no legitimate data of $I_{AB}$ above the $I_E$ curve any more. Here, we find $\eta_{\min}\simeq92.4\%$ and $85.8\%$ for symmetric and asymmetric Bell experiments, respectively.

Obviously, the curves in Fig.~\ref{fig1} match the analytic results derived in~\cite{ABG+07,PAB+09}; namely, the eavesdropping information
\begin{align}
I_E=h\left[ \frac{1\pm\sqrt{(2\mathsf{Q}+1)^2-1}}{2} \right],
\end{align}
and, with the QBER associated with $\mathsf{Q}$, the mutual information
\begin{align}
I_{AB}=1-h\left[\frac{1}{2}\left(1-\frac{2\mathsf{Q}+1}{\sqrt{2}}\right)\right].
\end{align}
The critical QBER and threshold efficiency can then be verified by letting $\mathcal{R}=0$. As mentioned in Introduction, if one adds one more output to register the undetected events, the input-output statistics with maximally entangled states can yield a slightly lower threshold efficiency down to approximately $90.8\%$~\cite{WAP21}, and via the convex-combination method~\cite{LB-JF+22} an efficient upper bound on the key rate implies a threshold efficiency as approximately $89.2\%$.

\subsubsection{The $I_{3322}$-based protocol}
For multi-input-two-output scenarios there is in general no nontrivial low-dimensional subspaces, the key rate evaluated with Bell-diagonal states is actually an estimated one. The $I_{3322}$ inequality reads $\mathsf{P}=P_{11}+P_{12}+P_{13}+P_{21}+P_{22}-P_{23}+P_{31}-P_{32}-2P_{10}-P_{20}-P_{01}\leq0$~\cite{CG04},
with the same short notations as in the CHSH inequality. Again, we pin down $I_E(\mathsf{Q})$ and $I_{AB}(\mathsf{Q})$, shown in Figs.~\ref{fig1}(e) through \ref{fig1}(h), with a very similar argument.

A distinct feature appears in the $I_{3322}$ scenario, however. Note that the cross point $(\mathsf{Q},I_E)=(\mathsf{Q},I_{AB})$ varies with respect to efficiency. For instance, in the asymmetric setup, one can get
\begin{equation}
  (\mathsf{Q},I_E)=(\mathsf{Q},I_{AB})\simeq(0.104, 0.678),
\end{equation}
as $\eta=1$, yielding $\varepsilon_{\rm cr}\simeq 5.86\%$; and
\begin{equation}
  (\mathsf{Q},I_E)=(\mathsf{Q},I_{AB})\simeq(0.127, 0.590),
\end{equation}
as $\eta\simeq83.6\%$, yielding $\varepsilon_{\rm cr}\simeq 8.22\%$. Along with an advantage over the threshold efficiency, i.e., $83.6\%$ here for the $I_{3322}$ case versus $85.8\%$ for the CHSH case, it is shown that in DIQKD protocols the $I_{3322}$ inequality seems more suitable than the CHSH inequality in the asymmetric setup, just as it indeed is in asymmetric BT~\cite{BGS+07}. It is henceforth implied that more measurements may indeed have distinct merits in asymmetric DIQKD.

Shown in Table~\ref{table2} are results with various other Bell inequalities. These inequalities reach quantum maxima with the maximally entangled state~\cite{BG08}. The explicit expressions of the inequalities are shown in Table~\ref{table3}. In~\cite{LB-JF+22}, a couple of multi-setting protocols proposed in~\cite{G-UPC21} are studied via the convex-combination method. Therein the threshold efficiencies are shown to be improved. Nevertheless, one protocol there involves a more-than-two-output Bell inequality and the other protocol there involves a Bell inequality whose violation requires high-dimensional quantum states. We have not studied the two protocols due to scope issues.

\begin{table}[b]
\caption{\label{table2}Estimated values of critical QBERs and efficiencies of protocols with various Bell inequalities in symmetric and asymmetric detection setups. The indices of the inequalities are similar to those in~\cite{BG08}.  }
\begin{ruledtabular}
\begin{tabular}{cccccc}
$\mathsf{P}$ & $\varepsilon_{\rm cr}~(\eta=1)$ & $\varepsilon_{\rm cr}~(\eta_{\min}^{\rm sym})$  & $\varepsilon_{\rm cr}~(\eta_{\min}^{\rm asym})$  & $\eta_{\min}^{\rm sym}$  & $\eta_{\min}^{\rm asym}$  \\
 \hline
CHSH  & 0.0715 & 0.0715 & 0.0715 & 0.924  &  {0.858}   \\
$I_{3322}$  & 0.0586 & 0.0586 & \fbox{0.0822} & 0.938  & \fbox{{0.836}}\\
$I_{4322}^{(3)}$  & 0.0586 & 0.0552  & 0.0586  &  0.942  &  {0.880}  \\
$A_6$  & 0.0532  & 0.0417  & 0.0426 &  0.957 & {0.916}  \\
$AS_1$ & 0.0663  & 0.0663  & 0.0663  & 0.927  & {0.864}  \\
$AS_2$ & 0.0698  & 0.0698  & 0.0698  & 0.924  & {0.858} \\
$AII_2$& 0.0679   & 0.0637   & 0.0537  & 0.932  & {0.893}  \\
$I_{4422}^{(5)}$& 0.0591   & 0.0559   & 0.0532  &  0.942 & {0.895}  \\
$I_{4422}^{(6)}$& 0.0532   &  0.0480  & 0.0433  & 0.950  & {0.915}  \\
$I_{4422}^{(13)}$& 0.0462   & 0.0501   & 0.0499  & 0.947  & {0.900}  \\
$I_{4422}^{(15)}$& 0.0473   & 0.0417   & 0.0428  & 0.955  & {0.915}  \\
$I_{4422}^{(19)}$& 0.0485   & 0.0485  & 0.0485  & 0.950  &  {0.905}
\end{tabular}
\end{ruledtabular}
\end{table}

\subsubsection{Reduction to the BB84 protocol}
\begin{figure}[b]
\subfigure[]{\includegraphics[width=40mm]{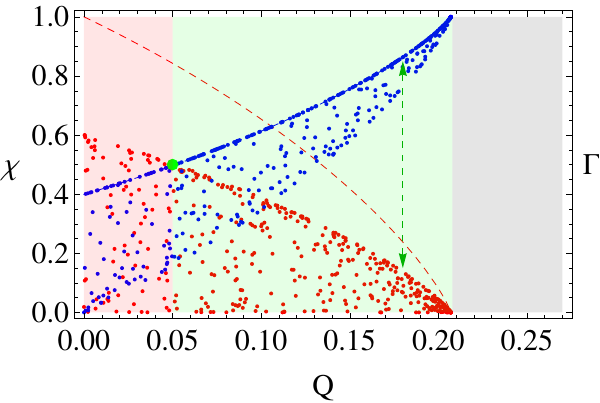}}\hspace{4mm}
\subfigure[]{\includegraphics[width=40mm]{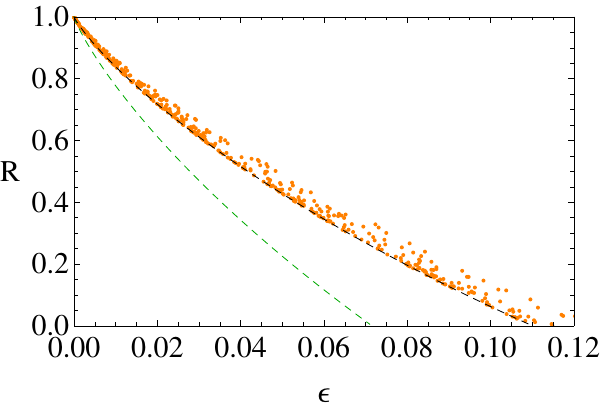}}\hspace{4mm}
  \caption{(Color online) (a) The $\chi$ and $\Gamma$ figures with respect to $\mathsf{Q}$ in the CHSH-based protocol with Alice's measurements specified. The red points represent the data of $(\mathsf{Q},\chi')$ (see the main text for the definition) from which to pin down the $I_E(\mathsf{Q})$. The blue points represent the data of $(\mathsf{Q},\Gamma)$ from which to pin down the $I_{AB}(\mathsf{Q})$. The dashed curves are analytic results of the BB84 protocol, plotted for the sake of comparison. (b) The $\mathcal{R}$ figure with respect to $\varepsilon$. The yellow points represent the date of $(\varepsilon,~\Gamma-\max\{\chi_1,\chi_2,\chi_3\})$. The (black) lower boundary represents the key rate of the BB84 protocol. The (green) curve denotes the key rate of the CHSH-based protocol. }\label{fig2}
\end{figure}

As the aforementioned, the DIQKD protocol will reduce to the BB84 protocol if one specifies Alice's measurements. In Fig.~\ref{fig2}, it is shown that in each step one randomly selects a Bell-diagonal state from which to get the $(\mathsf{Q},\Gamma)$ pair and the $(\mathsf{Q},\chi')$ pair, where $\chi'$ is attained by fixing Alice's measurements in evaluating $\rho_{E|a}$ along the $\hat z$-axis. The Monte Carlo procedure then yields the mutual and eavesdropping information as in the BB84 protocol. Note that the cross point $(\mathsf{Q},I_E)=(\mathsf{Q},I_{AB})\simeq(0.05,0.50)$ leads to the well-known critical QBER $\simeq11\%$.

\begin{table}[t]
\caption{\label{table3}Partial list of multi-setting Bell inequalities~\cite{BG08}.  }
\begin{ruledtabular}
\begin{tabular}{cp{160mm}}
 $\mathsf{P}$ & Expressions for each Bell inequality\\
 \hline
  CHSH & $P_{11}+P_{12}+P_{21}-P_{22}-P_{10}-P_{01}$\\
  $I_{3322}$ & $P_{11}+P_{12}+P_{13}+P_{21}+P_{22}-P_{23}+P_{31}-P_{32}-2P_{10}-P_{20}-P_{01}$\\
  $I_{4322}^{(3)}$ & $2 P_{11} + P_{12} + P_{13} - P_{21} + P_{22} + P_{23 }+ P_{32} - P_{33} + P_{41} - P_{42} - P_{43}- 2 P_{10}- P_{20} - P_{01} - P_{02}$ \\
  $A_6$ & $P_{11} + P_{12} + P_{14} + P_{21} + P_{23} - P_{24}+ P_{32} - P_{33} - P_{34} + P_{41} - P_{42}- P_{43}- P_{44}- P_{10} - P_{20} - P_{01} - P_{02}$  \\
  $AS_1$ & $P_{11} + P_{12} + P_{13} + P_{14} + P_{21} + P_{22} + P_{23} - P_{24} + P_{31} + P_{32} -2 P_{33}+ P_{41} - P_{42}- 2 P_{10} - P_{20}- 2 P_{01} - P_{02}$  \\
  $AS_2$ & $P_{11} + P_{12} + 2 P_{13} + 2 P_{14} + P_{21} + 2 P_{22}+ P_{23} - 2 P_{24} + 2 P_{31} + P_{32} -2 P_{33}+ P_{34}+ 2 P_{41} - 2 P_{42} + P_{43} - P_{44}- 3 P_{10} - P_{20} - P_{30} - 3 P_{01}- P_{02} - P_{03}$ \\
  $AII_2$ & $2 P_{11} + P_{12} + P_{13} - P_{14} + P_{21} + 2 P_{22} - P_{23} + P_{24} + P_{31} - P_{32 }- P_{33}+ P_{34} + P_{41} - P_{42}- P_{10} - P_{20}-3 P_{01} - P_{02} - P_{04}$\\
  $I_{4422}^{(5)}$ & $P_{11} + P_{13} + P_{21} + P_{22} - P_{23} + P_{24} + P_{31} - P_{32} + P_{41} + P_{42} - P_{43}- P_{44}- P_{10} - P_{20} - 2 P_{01} - P_{02}$\\
  $I_{4422}^{(6)}$ & $P_{11} - P_{12} + P_{13} + P_{14} + P_{21} + P_{22} - P_{23} + P_{24} + P_{31} - P_{32} + P_{33} - P_{34}+ P_{41} + P_{42} - P_{43} - P_{44}- P_{10} - P_{20} - 2 P_{01} - P_{02} - P_{03}$\\
  $I_{4422}^{(13)}$ & $P_{12} + P_{13} + P_{14} + P_{21} - 2 P_{22} + P_{23}+ P_{24} + P_{31} + P_{32} - P_{33} + P_{34}+ P_{41} + P_{42} + P_{43} - P_{44}- 2 P_{10}- P_{20} - P_{30} - 2 P_{01} - P_{02} - P_{03}$\\
  $I_{4422}^{(15)}$ & $2 P_{11} + P_{12} + P_{13} + P_{14} + P_{21} - P_{22}- P_{23} + P_{24} + P_{31} - P_{32} - P_{34}+P_{41} + P_{42} - P_{43} - P_{44}- 2 P_{10}- P_{20} - 2 P_{01} - P_{02}$\\
  $I_{4422}^{(19)}$ & $2 P_{11} + 2 P_{12} + P_{13} + 2 P_{14} + 2 P_{21}- P_{22} + 2 P_{23} - 2 P_{24} + P_{31} + 2 P_{32}- P_{33} - P_{34}+ 2 P_{41} - 2 P_{42} - P_{43}- 3 P_{10} - 2 P_{20} - 3 P_{01} - 2 P_{02}$
\end{tabular}
\end{ruledtabular}
\end{table}

\subsection{\label{beyond}Beyond $2\times2$ subspaces}
To have more precise estimates of the key rate, i.e., to lower the upper bounds obtained in $2\times2$ subspaces, one is required to consider arbitrarily high dimensional Hilbert spaces producing dichotomic outputs under unfaithful measurements with non-ideal efficiencies. It is then straightforward to generalize the Monte Carlo procedure in Table~\ref{table1} to make it work for high dimensions. The steps are similar as those in Table~\ref{table1}, except for a few quantities which must be adapted for high-dimensional states. The relevant equations are listed below.
\begin{align}
&\forall~d:~~\xi=1,2,...,d-1,~~\tau=u_i^{(\xi)},v_j^{(\xi)}\in SU(d),\label{proj:d-1}\\
&\hat p_\tau^{(0_\xi)}:=\sum_{t=0}^{\xi-1}\hat p_\tau^{(t)},~~~~\hat p_\tau^{(1_\xi)}:=\sum_{t=\xi}^{d-1}\hat p_\tau^{(t)},~~~~\Omega_\tau^{(\xi)}:=\hat p_\tau^{(0_\xi)}-\hat p_\tau^{(1_\xi)}=2\hat p_\tau^{(0_\xi)}-1,\label{proj:d-3}\\
&\mathsf{Q}^{(\xi)}=\max\Big\{ \mathsf{P}\Big| i,j\in M,~~u_i^{(\xi)},v_j^{(\xi)}\in SU(d) \Big\},\label{violation:xi}\\
&\varepsilon^{(\xi)}=\tr  \left[\left(\hat p_{\openone}^{(0_\xi)}\otimes\hat p_{\openone}^{(1_\xi)}+\hat p_{\openone}^{(1_\xi)}\otimes\hat p_{\openone}^{(0_\xi)} \right) \rho_{AB}\right],\label{errorrate:xi}\\
&\hat{p}^{(0_\xi)}_\tau\overset{\eta}{\longrightarrow}\eta\times\hat{p}^{(0_\xi)}_\tau+(1-\eta),~~~~\hat{p}^{(1_\xi)}_\tau\overset{\eta}{\longrightarrow}\eta\times\hat{p}^{(1_\xi)}_\tau,\\
&\varepsilon^{(\xi)}\longrightarrow\varepsilon^{(\xi)}(\eta_A,\eta_B),~~~~\Gamma^{(\xi)}=1-h\left(\varepsilon^{(\xi)}\right),\label{gamma:xi}\\
&\tilde{\rho}_{E|a}=\tr_{AB}\left[ \hat p^{(a)}_u\otimes\openone_{BE}\ketbra{\psi}{\psi}_{ABE} \right],~~~~\rho_{E|a}=\tilde{\rho}_{E|a}\big/\tr\tilde{\rho}_{E|a},\\
&\chi^{(\xi)}=S(\rho_E)-\sum_{a=0}^{d-1}q_a S(\rho_{E|a}),\label{eaves:holevo}\\
&\forall~\mathsf{Q}^{(\xi)}:~~I_E^{(\xi)}=\max\left\{ \chi^{(\xi)}\Big| \vec\zeta\in\mathbb{S}_{\mathsf{Q}^{(\vec\zeta)}}, ~u\in SU(d),~q_a\geq0,~\sum_{a=0}^{\xi-1}q_a=\sum_{a=\xi}^{d-1}q_a=1/2 \right\}:=I_E^{(\xi)}\left(\mathsf{Q}^{(\xi)}\right),\label{eaves:xi}\\
&~~~~~~~~~~~~~I_{AB}^{(\xi)}:=I_{AB}^{(\xi)}\left(\mathsf{Q}^{(\xi)}\right),\label{eaves:d}\\
&\mathcal{R}=\min\Big\{\min\Big\{ I_{AB}^{(\xi)}-I_E^{(\xi)} \Big| \xi=1,2,...,d-1\Big\}\Big|d=2,3,...,\infty\Big\}.\label{keyrate:xi}
\end{align}

We would like to elaborate now. Some steps in Table~\ref{table1} remain unchanged, except the following ones. For step 2, one needs to consider a high-dimensional state $\rho_{AB}=\sum_{i,j,m,n}C_{ij;mn}\ketbra{i}{j}_A\otimes\ketbra{m}{n}_B$. The coefficients $C_{ij;mn}$ must satisfy constraints such that the generated key, obtained by measurements in the basis of key generation, is random. A possible example could be $\rho_{AB}=\ketbra{\psi}{\psi}_{AB}$, with $\ket{\psi}_{AB}=\sum_{i=0}^{d-1}\frac{1}{\sqrt{d}}\ket{i}\otimes\ket{i}$. Due to the two-output feature of the protocol, some measuring results for the key should be denoted as 0 and the others as 1 (indexed by $\xi$ in (\ref{proj:d-1}); see the remarks below for detail). Other examples could be $\ket{\psi}_{AB}=\frac{1}{\sqrt{2}}(\ket{0}\otimes\ket{0}+\ket{d-1}\otimes\ket{d-1})$, $\ket{\psi}_{AB}=\frac{1}{\sqrt{2}}(\ket{0}\otimes\ket{0}+\ket{1}\otimes\ket{d-1})$, etc., or their convex combinations. The Holevo quantity can then be computed via (\ref{eaves:holevo}).

For step 3, to compute $\mathsf{Q}$ of a two-output Bell inequality $\mathsf{P}$, the projectors with respect to a certain $\xi$ are constructed in (\ref{proj:d-3}). With the above high-dimensional $\rho_{AB}$, one can have the quantum violation, i.e., (\ref{violation:xi}). For step 5, the eavesdropping information is attained in (\ref{eaves:xi}) by numerating all possible $q_a$'s. The constraints $\sum_{a=0}^{\xi-1}q_a=\sum_{a=\xi}^{d-1}q_a=1/2$ correspond to the requirement of randomness upon the key.

For steps 6 and 7, similarly, one considers a high-dimensional $\rho'_{AB}$ and compute the error rate in (\ref{errorrate:xi}) from which to get the mutual entropy under non-ideal efficiencies, i.e., (\ref{gamma:xi}). The quantum violation is computed with $\rho'_{AB}$. After collecting sufficient data, one can have the mutual information (\ref{eaves:d}) in step 9.

Finally, for step 10, one have attained a set of key rates (\ref{keyrate:xi}) with various $\xi$'s and $d$'s. An efficient upper bound of the key rate in the protocol is then taken as the minimum.

Further remarks are in order. First, equations~(\ref{proj:d-1})-(\ref{proj:d-3}) are defined so because the dimensions of the systems used should be presumed unknown while in measurements dichotomic results $\pm1$ should always be attained. One thus needs to numerate all possible divisions, here indicated by $\xi$'s, each of which denotes that the first $\xi$ dimensions yield result $0$ and the remaining dimensions yield result $1$ (or equivalently, through(\ref{proj:d-3}), the first $\xi$ dimensions yield result $+1$ and the remaining dimensions yield result $-1$). The divisions have included all possible reordering of results $\pm1$ since swap operations are among the unitary transformations. Bell inequalities of probabilistic form $\mathsf{P}$, because of (\ref{proj:d-3}), can be equivalent to inequalities of correlation form in which measurements of each party produce $\pm1$. For instance, the $I_{4422}^{(4)}$ inequality holds quantum mechanically with any $2\times2$-dimensional entangled states but can be violated with a $3\times3$- or $4\times4$-dimensional entangled state. Going beyond the $2\times2$-dimensional subspaces in the security analysis is henceforth very nontrivial (see also~\cite{G-UPC21}).

Second, it is similar, as with the $2\times2$ subspaces, to model undetected events and to compute the error rates and mutual entropy, which are all associated with outputs $\pm1$. In computing the Holevo quantity, however, it is essential to distinguish each dimension, in order to have the eavesdropping maximized whereby to pin down the $I_{AB}^{(\xi)}\left(\mathsf{Q}^{(\xi)}\right)$. In analogy to considering Bell-diagonal states $\mathbb{B}$ and $\mathbb{B}_\mathsf{Q}$ in the previous section, the $\mathbb{S}$ in (\ref{eaves:d}) represents a set of high-dimensional states shared between Alice and Bob and the $\mathbb{S}_{\mathsf{Q}^{(\vec\zeta)}}$ represents a subset of $\mathbb{S}$ that leads to some particular $\mathsf{Q}^{(\xi)}$ in the BT, with $\vec\zeta$ the parameters in the state in analogy to $\vec\Lambda$ in Bell-diagonal states. After pinning down a $\mathsf{Q}$-dependent $I_{AB}^{(\xi)}\left(\mathsf{Q}^{(\xi)}\right)$ in (\ref{eaves:d}), which follows the definition (\ref{mutual}), one can get a provisional upper bound on the key rate with respect to certain $\xi$ and $d$.

\section{Summary and outlook}
We have presented a Monte Carlo approach to evaluating information-theoretic security of DIQKD protocols. We have been able to asymptotically compute the key rates with the eavesdropping information and mutual information in terms of a same index, i.e., the $\mathsf{Q}$, via associating the QBER of the key with quantum violation of the Bell inequality. From the key rates, the critical QBERs and efficiencies of a family of protocols have been evaluated. The approach not only applies to scenarios with extremal efficiencies as in our case studies but also to those with general ones, and can subsequently be extended to multi-output protocols. Our results show that in either the symmetric or the asymmetric setup of the protocols not so many established multi-setting Bell inequalities, except for the $I_{3322}$ inequality so far among the family of inequalities we have considered, have better performance than the CHSH inequality. Hence it is a nontrivial task that one needs to select or build Bell inequalities for multi-setting protocols. The merits of the approach include: (i) the procedure is quite efficient, as it can reproduce the exact key rate in the CHSH-based protocol, while in general protocols it yields upper bounds on the key rates, and (ii) the asymmetric setups are closer to practical scenarios, since the nodes in a quantum network, for instance, usually have unequal detection efficiencies.

A couple of open questions remain. In our case studies, the eavesdropping information decreases as quantum violation increases; hence always the upper bound of mutual entropy should the mutual information correspond to. It remains an open question, then, of whether there are any counterexamples where $I_E(\mathsf{Q})$ takes reversed monotonicity; we conjecture there might be in protocols with more-than-two outputs. One more question is whether there are quantum algorithms to speed up the matrix diagonalization in evaluating the Holevo quantity for large $d$.

\acknowledgments
The author is grateful to Professor W.-Y. Hwang for enlightening the heralding method in DIQKD and BT, and the anonymous referees for giving very constructive comments. The study is supported by the National Natural Science Foundation of China (Grant No. 11905209) and the Fundamental Research Funds for the Central Universities (Grant No. 3072022TS2503).


\begin{thebibliography}{99}
\bibitem{Ekert91}
Ekert A K 1991
{Quantum cryptography based on Bell's Theorem}
\textit{Phys. Rev. Lett.} \textbf{67} 661

\bibitem{MY98}
Mayers D and Yao A
{Quantum cryptography with imperfect apparatus}
In \emph{Proceedings 39th Annual Symposium on Foundations of Computer Science} 503-509 (IEEE, 1998)

\bibitem{BHK05}
Barrett J, Hardy L and Kent A 2005
{No signaling and quantum key distribution}
\textit{Phys. Rev. Lett.} \textbf{95} 010503

\bibitem{AGM06}
Ac\'in A, Gisin N and Masanes L 2006
{From Bell's theorem to secure quantum key distribution}
\textit{Phys. Rev. Lett.} \textbf{97} 120405

\bibitem{AMP06}
Ac\'in A, Massar S and Pironio S 2006
{Efficient quantum key distribution secure against no-signalling eavesdroppers}
\textit{New J. Phys.} \textbf{8} 126

\bibitem{ABG+07}
Ac\'in A, Brunner N, Gisin N, Massar S, Pironio S and Scarani V 2007
{Device-independent security of quantum cryptography against collective attacks}
\textit{Phys. Rev. Lett.} \textbf{98} 230501

\bibitem{PAB+09}
Pironio S, Ac\'in A, Brunner N, Gisin N, Massar S and Scarani V 2009
{Device-independent quantum key distribution secure against collective attacks}
\textit{New J. Phys.} \textbf{11} 045021

\bibitem{PAM+10}
Pironio S, Ac\'in A,  Massar S, Boyer de la Giroday A, Matsukevich D N, Maunz P,  Olmschenk S, Hayes D, Luo L, Manning T A and Monroe C 2010
{Random numbers certified by Bell's theorem}
\textit{Nature} \textbf{464} 1021-1024

\bibitem{MPA11}
Masanes L, Pironio S and Ac\'in A 2011
{Secure device-independent quantum key distribution with causally independent measurement devices}
\textit{Nat. Commun.} \textbf{2} 1-7

\bibitem{AMP12}
Ac\'in A, Massar S and Pironio S 2012
{Randomness versus nonlocality and entanglement}
\textit{Phys. Rev. Lett.} \textbf{108} 100402

\bibitem{TH13}
Tomamichel M and H\"aggi E 2013
{The link between entropic uncertainty and nonlocality}
\textit{J. Phys. A: Math. Theor.} \textbf{46} 055301

\bibitem{VV14}
Vazirani U and Vidick T 2014
{Fully device-independent quantum key distribution}
\textit{Phys. Rev. Lett.} \textbf{113} 140501

\bibitem{MS16}
Miller C A and Shi Y 2016
{Robust protocols for securely expanding randomness and distributing keys using untrusted quantum devices}
\textit{J. ACM} \textbf{63} 1-63

\bibitem{A-FDF+18}
Arnon-Friedman R, Dupuis F, Fawzi O, Renner R and Vidick T 2018
{Practical device-independent quantum cryptography via entropy accumulation}
\textit{Nat. Commun.} \textbf{9} 1-11

\bibitem{A-FRV19}
Arnon-Friedman R, Renner R and Vidick T 2019
{Simple and tight device-independent security proofs}
\textit{SIAM J. Comput.} \textbf{48} 181-225

\bibitem{MvDR+19}
Murta G, van Dam S B, Ribeiro J, Hanson R and Wehner S 2019
{Towards a realization of device-independent quantum key distribution}
\textit{Quantum Sci. Technol.} \textbf{4} 035011

\bibitem{BCP+14}
Brunner N, Cavalcanti D, Pironio S, Scarani V and Wehner S 2014
{Bell nonlocality}
\textit{Rev. Mod. Phys.} \textbf{86} 419

\bibitem{HBD+15}
Hensen B, Bernien H, Dr\'eau A E, Reiserer A, Kalb N, Blok M S, Ruitenberg J, Vermeulen R F L, Schouten R N, Abell\'an C, Amaya W, Pruneri V, Mitchell M W, Markham M, Twitchen D J, Elkouss D, Wehner S, Taminiau T H and Hanson R 2015
{Loophole-free Bell inequality violation using electron spins separated by 1.3 kilometres}
\textit{Nature} \textbf{526} 682-686

\bibitem{GVW+15}
Giustina M, Versteegh M A M, Wengerowsky S, Handsteiner J, Hochrainer A, Phelan K, Steinlechner F, Kofler J, Larsson J-A, Abell\'an C, Amaya W, Pruneri V, Mitchell M W, Beyer J, Gerrits T, Lita A E, Shalm L K, Nam S W, Scheidl T, Ursin R, Wittmann B and Zeilinger A 2015
{Significant-loophole-free test of Bell's theorem with entangled photons}
\textit{Phys. Rev. Lett.} \textbf{115} 250401

\bibitem{SM-SC+15}
Shalm L K, Giustina M, Meyer-Scott E, Christensen B G, Bierhorst P, Wayne M A, Stevens M J, Gerrits T, Glancy S, Hamel D R, Allman M S, Coakley K J, Dyer S D, Hodge C, Lita A E, Verma V B, Lambrocco C, Tortorici E, Migdall A L, Zhang Y, Kumor D R, Farr W H, Marsili F, Shaw M D, Stern J A, Abell\'an C, Amaya W, Pruneri V, Jennewein T, Mitchell M W, Kwiat P G, Bienfang J C, Mirin R P, Knill E and Nam S W 2015
{Strong loophole-free test of local realism}
\textit{Phys. Rev. Lett.} \textbf{115} 250402

\bibitem{RBG+17}
Rosenfeld W, Burchardt D, Garthoff R, Redeker K, Ortegel N, Rau M and Weinfurter H 2017
{Event-ready Bell test using entangled atoms simultaneously closing detection and locality loopholes}
\textit{Phys. Rev. Lett.} \textbf{119} 010402

\bibitem{LWZ+18}
Li M-H, Wu C, Zhang Y, Liu W-Z, Bai B, Liu Y, Zhang W, Zhao Q, Li H, Wang Z, You L, Munro W J, Yin J, Zhang J, Peng C-Z, Ma X, Zhang Q, Fan J and Pan J-W 2018
{Test of local realism into the past without detection and locality loopholes}
\textit{Phys. Rev. Lett.} \textbf{121} 080404

\bibitem{Bell04}
Bell J S
In \emph{Speakable and unspeakable in quantum mechanics} (Cambridge University Press, 2004) chap. 24.

\bibitem{Pearle70}
Pearle P M 1970
{Hidden-variable example based upon data rejection}
\textit{Phys. Rev. D.} \textbf{2} 1418

\bibitem{CH74}
Clauser J F and Horne M A 1974
{Experimental consequences of objective local theories}
\textit{Phys. Rev. D.} \textbf{10} 526

\bibitem{GG99}
Gisin N and Gisin B 1999
{A local hidden variable model of quantum correlation exploiting the detection loophole}
\textit{Phys. Lett. A} \textbf{260} 323-327

\bibitem{Massar02}
Massar S 2002
{Nonlocality, closing the detection loophole, and communication complexity}
\textit{Phys. Rev. A} \textbf{65} 032121

\bibitem{MPR+02}
Massar S, Pironio S, Roland J and Gisin B 2002
{Bell inequalities resistant to detector inefficiency}
\textit{Phys. Rev. A} \textbf{66} 052112

\bibitem{MP03}
Massar S and Pironio S 2003
{Violation of local realism versus detection efficiency}
\textit{Phys. Rev. A} \textbf{68} 062109

\bibitem{BG08}
Brunner N and Gisin N 2008
{Partial list of bipartite Bell inequalities with four binary settings}
\textit{Phys. Lett. A} \textbf{372} 3162-3167

\bibitem{PV09}
P\'al K F and V\'ertesi T 2009
{Quantum bounds on Bell inequalities}
\textit{Phys. Rev. A} \textbf{79} 022120

\bibitem{VPB10}
V\'ertesi T, Pironio S and Brunner N 2010
{Closing the Detection Loophole in Bell Experiments Using Qudits}
\textit{Phys. Rev. Lett.} \textbf{104} 060401

\bibitem{Branciard11}
Branciard C 2011
{Detection loophole in Bell experiments: How postselection modifies the requirements to observe nonlocality}
\textit{Phys. Rev. A} \textbf{83} 032123

\bibitem{Bell64}
Bell J S 1964
{On the Einstein-Podolsky-Rosen paradox}
\textit{Phys. Phys. Fiz.} \textbf{1} 195

\bibitem{CHS+69}
Clauser J F, Horne M A, Shimony A and Holt R A 1969
{Proposed experiment to test local hidden-variable theories}
\textit{Phys. Rev. Lett.} \textbf{23} 880

\bibitem{ZLA-F+23}
Zapatero V, van Leent T, Arnon-Friedman R, Liu W-Z, Zhang Q, Weinfurter H and Curty M 2023
{Advances in device-independent quantum key distribution}
\textit{npj Quantum Inf.} \textbf{9} 10

\bibitem{XMZ+20}
Xu F, Ma X, Zhang Q, Lo H-K and Pan J-W 2020
{Secure quantum key distribution with realistic devices}
\textit{Rev. Mod. Phys.} \textbf{92} 025002

\bibitem{PR22}
Portmann C and Renner R 2022
{Security in quantum cryptography}.
\textit{Rev. Mod. Phys.} \textbf{94} 025008

\bibitem{PGT+23}
Primaatmaja I W, Goh K T, Tan E Y-Z,  Khoo J T-F, Ghorai S and Lim C C-W 2023
{Security of device-independent quantum key distribution protocols: a review}
\textit{Quantum} \textbf{7} 932

\bibitem{NFL+21}
Niemietz D, Farrera P, Langenfeld S and Rempe G 2021
{Nondestructive detection of photonic qubits}
\textit{Nature} \textbf{591} 570-574

\bibitem{ZC19}
Zapatero V and Curty M 2019
{Long-distance device-independent quantum key distribution}
\textit{Sci. Rep.} \textbf{9} 1-18

\bibitem{GPS10}
Gisin N, Pironio S and Sangouard N 2010
{Proposal for implementing deviceindependent quantum key distribution based on a heralded qubit amplifier}
\textit{Phys. Rev. Lett.} \textbf{105} 070501

\bibitem{PMW+11}
Pitkanen D, Ma X, Wickert R, van Loock P and L\"utkenhaus N 2011
{Efficient heralding of photonic qubits with applications to device-independent quantum key distribution}
\textit{Phys. Rev. A} \textbf{84} 022325

\bibitem{CM11}
Curty M and Moroder T 2011
{Heralded-qubit amplifiers for practical deviceindependent quantum key distribution}
\textit{Phys. Rev. A} \textbf{84} 010304R

\bibitem{M-SBB+13}
Meyer-Scott E, Bula M, Bartkiewicz K, \v{C}ernoch A, Soubusta J, Jennewein T and Lemr K 2013
{Entanglement-based linear-optical qubit amplifier}
\textit{Phys. Rev. A} \textbf{88} 012327

\bibitem{KMS+20}
Ko{\l}ody\'nski J, M\'attar A, Skrzypczyk P, Woodhead E, Cavalcanti D, Banaszek K and Ac\'in A 2020
{Device-independent quantum key distribution with singlephoton sources}
\textit{Quantum} \textbf{4} 260

\bibitem{ML12}
Ma X and L\"ukenhaus N 2012
{Improved Data Post-Processing in Quantum Key Distribution and Application to Loss Thresholds in Device Independent QKD}
\textit{Quantum Inf. Comput.} \textbf{12} 0203-0214

\bibitem{BGS+07}
Brunner N, Gisin N, Scarani V and Simon C 2007
{Detection Loophole in Asymmetric Bell Experiments}
\textit{Phys. Rev. Lett.} \textbf{98} 220403

\bibitem{BMD+04}
Blinov B B, Moehring D L, Duan L-M and Monroe C 2004
{Observation of entanglement between a single trapped atom and a single photon}
\textit{Nature} \textbf{428} 153

\bibitem{MMB+04}
Moehring D L, Madsen M J, Blinov B B and Monroe C 2004
{Experimental Bell Inequality Violation with an Atom and a Photon}
\textit{Phys. Rev. Lett.} \textbf{93} 090410

\bibitem{VWS+06}
Volz J, Weber M, Schlenk D, Rosenfeld W, Vrana J, Saucke K, Kurtsiefer C and Weinfurter H 2006
{Observation of Entanglement of a Single Photon with a Trapped Atom}
\textit{Phys. Rev. Lett.} \textbf{96} 030404

\bibitem{BFF21}
Brown P, Fawzi H and Fawzi O 2021
{Device-independent lower bounds on the conditional von Neumann entropy}
\textit{arXiv:2106.13692}

\bibitem{WAP21}
Woodhead E, Ac\'in A and Pironio S 2021
{Device-independent quantum key distribution with asymmetric CHSH inequalities}
\textit{Quantum} \textbf{5} 443

\bibitem{BFF21-2}
Brown P, Fawzi H and Fawzi O 2021
{Computing conditional entropies for quantum correlations}
\textit{Nature Communications} \textbf{12} 575

\bibitem{MPW22}
Masini M, Pironio S and Woodhead E 2022
{Simple and practical DIQKD security analysis via BB84-type uncertainty relations and Pauli correlation constraints}
\textit{Quantum} \textbf{6} 843

\bibitem{LB-JF+22}
{\L}ukanowski K, Balanz\'{o}-Juand\'{o} M, Farkas M, Ac\'{i}n A and Ko{\l}ody\'{n}ski J 2022
Upper bounds on key rates in device-independent quantum key distribution based on convex-combination attacks
\textit{arXiv:2206.06245}

\bibitem{CG04}
Collins D and Gisin N 2004
{A relevant two qubit Bell inequality inequivalent to the CHSH inequality}
\textit{J. Phys. A: Math. Theor.} \textbf{37} 1775

\bibitem{BB84}
Bennett C H and Brassard G
{Quantum cryptography: Public key distribution and coin tossing}.
in \emph{Proceedings IEEE International Conference on Computers, Systems and Signal Processing, Bangalore, India, 1984} (IEEE, New York, 1984), pp. 175-179

\bibitem{ZZH97}
Zukowski M, Zeilinger A and Horne M A 1997
{Realizable higher-dimensional two-particle entanglements via multiport beam splitters}
\textit{Phys. Rev. A} \textbf{55} 2564

\bibitem{G-UPC21}
Gonzales-Ureta J R, Predojevi\'c A and Cabello A 2021
{Device-independent quantum key distribution based on Bell inequalities with more than two inputs and two outputs}
\textit{Phys. Rev. A} \textbf{103} 052436

\bibitem{DW05}
Devetak I and Winter A 2005
{Distillation of secret key and entanglement from quantum states}
\textit{Proc. R. Soc. A} \textbf{461} 207-235

\bibitem{Su22}
Su H-Y 2022
{A simple relation of guessing probability in quantum key distribution}
\textit{New J. Phys.} \textbf{24} 093016

\end{thebibliography}
\end{document}